\documentclass[12pt,preprint]{aastex}
\def\gsim{\ \raise 3pt \hbox{$>$} \kern -8.5pt \raise -2pt \hbox{$\sim$}\ }
\def\lsim{\ \raise 3pt \hbox{$<$} \kern -8.5pt \raise -2pt \hbox{$\sim$}\ }
\begin{document}
\title{Diffusive Synchrotron Radiation from Relativistic Shocks  in Gamma-Ray
Burst Sources}
\author{Gregory D. Fleishman\altaffilmark{1} }

\affil{ $^1$National Radio Astronomy Observatory, Charlottesville,
VA 22903 }

\begin{abstract}
The spectrum of electromagnetic emission generated by relativistic
electrons scattered on small-scale random magnetic fields, implied
by current models of the magnetic field generation in the
gamma-ray burst sources, is considered. The theory developed
includes both perturbative and non-perturbative versions and,
therefore, suggests a general treatment of the radiation in
arbitrary small-scale random field. It is shown that a general
treatment of the random nature of the small-scale magnetic field,
as well as angular diffusion  of the electrons due to multiple
scattering by magnetic inhomogeneities (i.e., non-perturbative
effects), give rise to a radiation spectrum that differs
significantly from so-called ``jitter" spectrum. The spectrum of
diffusive synchrotron radiation   seems to be consistent with the
low energy spectral index distribution of the gamma-ray bursts.

\end{abstract}

\section{Introduction}

Relativistic objects including, but not limited to, Gamma-Ray
Burst (GRB) sources are known to produce nonthermal emission,
which typically requires the generation of extra magnetic field,
at least at some stage of ejecta expansion after initial prompt
energy release \citep{Meszaros_2002}. A number of recent
publications
\citep{Kazimura_1998,Medvedev_Loeb_1999,Nishikawa_etal_2003,Jaroshek_etal_2004,Jaroshek_etal_2005}
put forward the concept of  magnetic field generation by a
relativistic version of the Weibel instability
\citep{Weibel_1959}. Analytical considerations
\citep{Medvedev_Loeb_1999}, together with more sophisticated
numerical modelling
\citep{Nishikawa_etal_2003,Nishikawa_etal_2005,Jaroshek_etal_2004,Jaroshek_etal_2005,Hededal_Nishikawa_2005},
suggest that the magnetic field produced by the Weibel instability
is random and extremely small-scale, with a characteristic
correlation length $l_{cor} \sim (0.1-1)l_{sk}$,  where
$l_{sk}=c/\omega_{pe}$ is the plasma skin depth, $\omega_{pe}$ is
the electron plasma frequency, and $c$ is the speed of light.

Depending on the magnetic field saturation value $B$, this
correlation length may be less than the coherence length (or
formation zone) of synchrotron radiation:
\begin{equation}
\label{Syn_coh_length}
  l_s=\frac{mc^2}{eB}=\frac{c}{\omega_{Be}}.
\end{equation}
Here, $e$ and $m$ are the electron
charge and mass, and $\omega_{Be}=\frac{eB}{mc}$ is the electron
gyrofrequency. Generation of electromagnetic emission by
relativistic electrons moving in small-scale magnetic fields is
known to differ from the case of emission by electrons in a
uniform (large-scale) magnetic field \citep{LL_2}.

Electromagnetic radiation in small-scale \emph{random} magnetic
fields was first calculated by \cite{Nik_Tsyt_1979} using
perturbation theory, where the trajectory of the particle is
assumed to be rectilinear, while the acceleration produced by the
random variations of the Lorentz force is taken into account by
the first non-zero order of the perturbation theory. The full
non-perturbative theory was later worked out by
\cite{Topt_Fl_1987a,Topt_etal_1987}, allowing in particular net
deflections of the radiating particles that are not necessarily
small (e.g., by  a large-scale magnetic field and/or multiple
scattering on small-scale inhomogeneities).

Recently, interest in this emission process in the context of
gamma-ray bursts was revived by \cite{Medvedev_2000}, who
estimated radiation spectrum in the presence of the small-scale
magnetic field and referred to the resulting radiation as ``jitter
radiation". An oversimplification of the estimates performed by
\cite{Medvedev_2000} resulted in the low-frequency asymptotic
limit of the jitter spectrum, $P(\omega) \propto \omega$, which is
applicable to  a simpler (undulator-like) version of what is
considered here, while may noticeably deviate from the correct
spectral shape in a general case. In this paper  a general
treatment of the radiation produced by relativistic electrons
moving through small-scale fluctuating magnetic fields, probably
available at the relativistic shocks, is presented. For the
completeness and convenience of further applications, we first
formulate a simplified perturbative version of the theory, and
then specify its region of applicability by comparing with the
available non-perturbative version. Possible implications of the
general theory for GRB physics are discussed.

\section{Perturbation theory of diffusive synchrotron radiation}

Let us calculate the radiation by a single relativistic particle
moving in a random small-scale magnetic field  based on the
equation for the radiated energy resulting from perturbation
theory \citep{LL_2}

\begin{equation}
\label{cal_E_w_perp}
  \frac{dW_{\omega}}{d\omega}=\frac{2\pi e^2}{c^3} \int_{1/2\gamma_*^2}^{\infty}
   \left(\frac{\omega}{\omega
  '}\right)^2 \left|{\bf w}_{\omega ' \bot}\right|^2
  \left[1-\frac{\omega}{\omega'\gamma_*^2} +
  \frac{\omega^2}{2\omega'^2\gamma_*^4  }
\right]
   d\left(\frac{\omega'}{\omega}\right).
\end{equation}

Here $\gamma_*=\gamma=E/mc^2$ in the vacuum, while
$\gamma_*^{-2}=\left(\gamma^{-2}+\frac{\omega_{pe}^2}{\omega^2}\right)$
in the presence of the background plasma,
$\omega'=\frac{\omega}{2}(\gamma^{-2}+\theta^2 +
\omega_{pe}^2/\omega^2)$, $\theta$ is the angle between the
electron speed and the direction of the emission, ${\bf w}_{\omega
' \bot}$ is the Fourier transform of the particle acceleration
${\bf w}_{\bot}={\bf F}/(m\gamma)$ transverse to the particle
speed, and ${\bf F}$ is the corresponding transverse component of
the Lorentz force. Evidently, for a random Lorentz force the
acceleration is a random function as well, although the average of
$\left|{\bf w}_{\omega ' \bot}\right|^2$ includes a regular (not
random) component, which eventually specifies the emission
pattern.

To find the averaged value of $\left|{\bf w}_{\omega '
\bot}\right|^2$, assume that the random Lorentz force (${\bf
F}({\bf r},t)$, which can include either magnetic and/or electric
fields) perturbing the rectilinear particle motion, can be
expressed as a Fourier integral over $\omega$ and ${\bf k}$:
\begin{equation}
\label{E_four}
    {\bf F}({\bf r},t)= \int e^{-i(\omega t - {\bf kr})}{\bf F}_{\omega, {\bf
k}}
    d\omega d{\bf k}.
\end{equation}
This force represents a global field, which can vary in space and
time in general. However, the particle acceleration is produced by
a local value of this field related to instantaneous positions of
the particle along its actual trajectory. Within the perturbation
theory we can adopt ${\bf r}= {\bf r}_0 + {\bf v}t$ for the
independent argument of the force acting on  the moving particle:
\begin{equation}
\label{E_appr}
    {\bf F}({\bf r}_0+{\bf v}t,t)= \int e^{-i(\omega t - {\bf kr}_0- {\bf kv}t)}
    {\bf F}_{\omega, {\bf k}} d\omega d{\bf k}.
\end{equation}
Then the Fourier component ${\bf F}_{\omega'}$, specifying the
magnitude ${\bf w}_{\omega'}={\bf
F}_{\omega'}/{m\gamma}$, can be found by direct temporal Fourier
transform of (\ref{E_appr}):
\begin{equation}
\label{E_four_1}
 {\bf F}_{\omega'}=\int\frac{dt}{2\pi} e^{i\omega't} {\bf F}({\bf r}_0+{\bf
v}t,t)=
 \int d\omega d{\bf k} \delta(\omega'-\omega+{\bf kv}) {\bf F}_{\omega, {\bf k}}
    e^{i{\bf kr}_0}.
\end{equation}
Now it is straightforward to write down the square of the force
modulus $\mid {\bf F}_{\omega'} \mid^2$:
\begin{equation}
\label{E_four_2}
 \mid {\bf F}_{\omega'} \mid^2=
 \int d\omega d{\bf k} d\omega_1 d{\bf k}_1
 e^{i({\bf k}-{\bf k}_1){\bf r}_0}
 \delta(\omega'-\omega+{\bf kv}) \delta(\omega'-\omega_1+{\bf k}_1{\bf v})
 {\bf F}_{\omega, {\bf k}}{\bf F}_{\omega_1, {\bf k}_1}^*.
\end{equation}

This value is still fluctuating in the random field, although its
mean value is non-zero (in contrast to the fluctuating field
itself ${\bf F}_{\omega, {\bf k}}$, which is  a complex random
function with zero mean). A steady-state level of radiation is
evidently specified by the average value $\left< \mid {\bf
F}_{\omega'} \mid^2 \right>$ of (\ref{E_four_2}). The easiest way
to perform the averaging is to average Eq. (\ref{E_four_2}) over
all possible initial positions ${\bf r}_0$ of the particle inside
the source. In practice, this corresponds to a uniform
distribution of radiating particles within a volume containing the
field inhomogeneities, which have a statistically uniform
distribution within the same volume. Since the exponent
$\exp({i({\bf k}-{\bf k}_1){\bf r}_0})$ oscillates strongly for
${\bf k}\neq{\bf k}_1$, we obtain:
\begin{equation}
\label{stat_unif}
 \frac{1}{V}
   \int e^{i({\bf k}-{\bf k}_1){\bf r}_0} d{\bf r}_0=
   \frac{(2\pi)^3}{V}\delta({\bf k}-{\bf k}_1),
 \end{equation}
where $V$ is the source volume, and, for $\left< \mid {\bf
F}_{\omega'} \mid^2 \right>$ obtain accordingly
\begin{equation}
\label{E_four_3}
 \left< \mid {\bf F}_{\omega'} \mid^2 \right>=\frac{(2\pi)^3}{V}
 \int  d\omega d{\bf k}
  \delta(\omega'-\omega+{\bf kv}) \mid
 {\bf F}_{\omega, {\bf k}} \mid^2
    .
\end{equation}
Then the mean value of the Fourier transform of the particle
acceleration is expressed as
\begin{equation}
\label{w_perp_3}
 \mid {\bf w}_{\omega'\bot} \mid^2=\frac{(2\pi)^3}{m^2 \gamma^2 V}
 \int  d\omega d{\bf k}
  \delta(\omega'-\omega+{\bf kv}) \mid {\bf F}_{\omega, {\bf k}}
  \mid^2,
\end{equation}
which provides a unique correspondence between the \emph{temporal
and spatial} Fourier transform of the Lorentz force (right-hand
side) and the \emph{temporal} Fourier transform of the particle
acceleration (left-hand side). Substituting (\ref{w_perp_3}) into
(\ref{cal_E_w_perp}), using dummy variables ($q_0, {\bf q}$) for
($\omega, {\bf k}$), we obtain finally for the radiated energy per
unit range of $\omega$
\begin{equation}
\label{cal_E_w_perp_3}
  \frac{dW_{\omega}}{d\omega}=\frac{(2\pi)^4e^2}{m^2c^3\gamma^2V}
  \int_{1/2\gamma_*^2}^{\infty} d\left(
  \frac{\omega'}{\omega} \right)
  \left(\frac{\omega}{\omega '}\right)^2
  \left[1-\frac{\omega}{\omega'\gamma_*^2} +
  \frac{\omega^2}{2\omega'^2\gamma_*^4  }
\right]
    \int  dq_0 d{\bf q}
  \delta(\omega'-q_0+{\bf qv}) \mid
 {\bf F}_{q_0, {\bf q}} \mid^2.
\end{equation}

The spectrum of the radiation described by Eq.
(\ref{cal_E_w_perp_3}) depends on the statistical properties of
the random force. To be more specific let us consider a random
magnetic field, i.e., $\mid {\bf F}_{q_0, {\bf q}} \mid^2=e^2\mid
{\bf B}^{\bot}_{q_0, {\bf q}} \mid^2 = e^2 (\delta_{\alpha\beta} -
v_{\alpha}v_{\beta}/v^2) B^{\alpha}_{q_0, {\bf q}}B^{\beta
*}_{q_0, {\bf q}}$, and introduce the second-order correlation
tensor of the statistically uniform random magnetic field as
follows \citep{Toptygin_83}:
\begin{equation}
\label{Corr_2} K_{\alpha \beta}^{(2)}({\bf r},\tau) = \left<B_{st,
\alpha}({\bf R},t)B_{st,\beta}({\bf R}+{\bf r},t+\tau)\right>
=\frac{1}{TV}\int dtd{\bf R} B_{st, \alpha}({\bf
R},t)B_{st,\beta}({\bf R}+{\bf r},t+\tau).
\end{equation}
Then, express $\mid {\bf B}^{\bot}_{q_0, {\bf q}} \mid^2$ via the
Fourier spectrum $K_{\alpha \beta}^{(2)}(q_0,{\bf q})$ of this
correlation tensor\footnote{Indeed, $B^{\alpha}_{q_0, {\bf
q}}B^{\beta *}_{q_0, {\bf q}} \equiv
 \int \frac{dt'dt''d{\bf r}'d{\bf
r}''}{(2\pi)^{8}}\exp[i(q_0t'-{\bf qr}')-i(q_0t''-{\bf qr}'')]
B_{st, \alpha}({\bf r}',t')B_{st,\beta}({\bf r}'',t'') =
 \int
\frac{d\tau d{\bf r}}{(2\pi)^{8}}\exp[i(q_0\tau-{\bf qr})] \int
dt' d{\bf r}' B_{st, \alpha}({\bf r}',t')B_{st,\beta}({\bf
r}'+{\bf r},t'+\tau)) =
 \frac{TV}{(2\pi)^4} \int \frac{d\tau d{\bf
r}}{(2\pi)^{4}}\exp[i(q_0\tau-{\bf qr})]K_{\alpha
\beta}^{(2)}({\bf r},\tau)=
 \frac{TV}{(2\pi)^4} K_{\alpha
\beta}^{(2)}(q_0,{\bf q})$. Here we used the correlation tensor
defined by Eq. (\ref{Corr_2}), whose Fourier transform is $
K_{\alpha\beta}^{(2)}(q_0,{\bf q})= \int {d{\bf r} d \tau \over (2
\pi)^4} e^{i(q_0 \tau - {\bf qr})} K_{\alpha \beta}^{(2)}({\bf
r},\tau)$. Then, multiplying this by $(\delta_{\alpha\beta} -
v_{\alpha}v_{\beta}/v^2)$, obtain (\ref{B_corr}).}:

\begin{equation}
\label{B_corr}
       \mid {\bf B}^{\bot}_{q_0,{\bf q}} \mid^2
   = \frac{TV}{(2\pi)^4}\left(\delta_{\alpha\beta} -
\frac{v_{\alpha}v_{\beta}}{v^2}\right) K_{\alpha
\beta}^{(2)}(q_0,{\bf q})
   = \frac{TV}{(2\pi)^4}
   K(q_0,{\bf q}) %\delta(q_0-q_0({\bf q}))
\end{equation}
where $K(q_0,{\bf q})=K_{\alpha\beta}^{(2)}(q_0,{\bf
q})\left(\delta_{\alpha\beta} -
\frac{v_{\alpha}v_{\beta}}{v^2}\right)$ is the spectrum of the
random magnetic field transverse to the particle velocity. If the
random field is composed of random waves with the dispersion
relation $q_0=q_0({\bf q})$ (in the limiting case of a static
random field we have $q_0({\bf q}) \equiv 0$), the spectrum takes
the form $K(q_0,{\bf q})= K({\bf q})\delta(q_0-q_0({\bf q}))$.

Substituting all required values into (\ref{w_perp_3}) and
dividing by the total duration of emission $T$ we arrive at the
radiation intensity (energy emitted per unit frequency per unit
time):
\begin{equation}
\label{I_perp_1}
   \frac{dI_{\omega}}{d\omega}=\frac{e^4}{m^2c^3\gamma^2}
  \int_{1/2\gamma_*^2}^{\infty} d\left(
  \frac{\omega'}{\omega} \right)
  \left(\frac{\omega}{\omega '}\right)^2
  \left[1-\frac{\omega}{\omega'\gamma_*^2} +
  \frac{\omega^2}{2\omega'^2\gamma_*^4  }
\right]
    \int  dq_0 d{\bf q}
  \delta(\omega'-q_0+{\bf qv}) K({\bf q})\delta(q_0-q_0({\bf q})),
\end{equation}
which eventually results in radiation spectra different from one
estimated by \cite{Medvedev_2000}.

\section{Illustrative examples}

It is now straightforward to consider a few different models of
the random magnetic field. Let us  proceed with the case of
quasi-static magnetic inhomogeneities, $K_{\alpha
\beta}^{(2)}(q_0,{\bf q}) = K_{\alpha \beta}^{(2)}({\bf
q})\delta(q_0)$, which is applicable for magnetic inhomogeneities
moving slower than the radiating particle. In the general case we
assume:
\begin{equation}
\label{power_spectr_2}
 K_{\alpha
\beta}^{(2)}({\bf q})=C_{\alpha \beta}\left<B_{st}^2\right>f({\bf
q}),
 \end{equation}
where the tensor structure of $C_{\alpha \beta}$ should be
consistent with Maxwell's Equation $\nabla \cdot {\bf B}=0$ and
provide the correct normalization $\int K_{\alpha
\alpha}^{(2)}({\bf q})d{\bf q}=\left<B_{st}^2\right>$.

Consider  the isotropic case $
 K_{\alpha
\beta}^{(2)}({\bf q}) \propto \left(\delta_{\alpha\beta} -
q_{\alpha}q_{\beta}/q^2\right)f(\mid {\bf q} \mid) $ with  a
specific power-law for the spectrum of random magnetic field:
\begin{equation}
\label{power_law_spectr}
 f({\bf q})=
 \frac{q^2}{(q_{m}^2+q^2)^{\nu/2+2}},
\end{equation}
where $\nu$ is the high-frequency spectral index of the magnetic
turbulence. The integral over $d{\bf q}$ then takes the form:
\begin{equation}
\label{Int_q_2}
       \int  d{\bf q}
   \frac{q^2\delta(\omega'+{\bf qv})}{(q_{m}^2+q^2)^{\nu/2+2}}=\frac{2\pi}{v}
  \left[\frac{1}{\nu}\left(\left(\frac{\omega'}{v}\right)^2+q_{m}^2\right)^{-
\nu/2}-
  \frac{q_{m}^2}{\nu+2}\left(\left(\frac{\omega'}{v}\right)^2+q_{m}^2\right)^{-
\nu/2-1}\right].
\end{equation}

Subsequent integration over $d(\omega'/\omega)$ yields the
radiation spectrum over the entire spectral range. We consider the
low-frequency and high-frequency asymptotic behavior of the
spectrum. At high frequencies, $\omega \gg \omega_0 \gamma^2$,
where $\omega_0=q_mc$, $q_m$ can be discarded everywhere in Eq.
(\ref{Int_q_2}), so after integration over $d(\omega'/\omega)$ we
have $dI_{\omega}/d\omega \propto \omega^{-\nu}$ and the radiation
spectrum resembles the turbulence spectrum at $q \gg q_m$. At low
frequencies, $\omega \ll \omega_0 \gamma^2$, the term $\omega'/v$
can be discarded in (\ref{Int_q_2}), which leads to the
frequency-independent part of the spectrum $dI_{\omega}/d\omega
\propto \omega^0$. At even lower frequencies,
$\omega<\omega_{pe}\gamma$, when the difference between $\gamma$
and $\gamma_*$ is important, the term $\omega'/v$ dominates again
giving rise to the spectral asymptote $dI_{\omega}/d\omega \propto
\omega^{2}$.

Note that this property does not depend on the specific shape of
the correlation function  $f({\bf q})$ in the range of small $q
\ll q_m$ ($\propto q^2$ in our case), and remains valid for any
shape peaking at $q_m$. To see this explicitly, consider the
following simplified form of the spectrum: $f({\bf q}) \propto
q^{\alpha}$ at $q < q_m$ and $f({\bf q}) = 0$ at $q > q_m$.
Therefore, in place of (\ref{Int_q_2}) we have:
\begin{equation}
\label{Int_q_2a}
       \int  d{\bf q} q^{\alpha}\delta(\omega'+{\bf qv})=\frac{2\pi}{(\alpha+2)v}
  \left[q_{m}^{\alpha+2}
  -\left(\frac{\omega'}{v}\right)^{\alpha+2}\right],
\end{equation}
which does not depend on $\omega$ at  $\omega' \ll \omega_0$ (or
$\omega \ll \omega_0  \gamma^2$) for $\alpha > -2$, thus, for any
fluctuation \emph{energy} spectrum, $q^2f({\bf q})$, peaking at
$q_m$. In particular, this is also valid for the Bremsstrahlung
spectrum, which arises as the fast particle experiences random
accelerations at microscopic scales by randomly distributed
Coulomb centers.

We now demonstrate that the same is valid for the two-dimensional
geometry implied by the current models of magnetic field
generation at the relativistic shocks \citep{Medvedev_Loeb_1999,
Nishikawa_etal_2003,Jaroshek_etal_2004,Jaroshek_etal_2005}.
Consider a specific case, when the observer looks along the $x$
axis, the shock front lies in the $xy$ plane, and the random field
belongs to the shock front, first examined by
\cite{Medvedev_2000}, who obtained a result different from one
given below. Note, that now the most general form of $K_{\alpha
\beta}({\bf q})$ is
\begin{equation}
\label{f_model_1}
 K_{\alpha \beta}^{(2)}({\bf q}) \propto \left(\delta_{\alpha\beta} -
n_{\alpha}n_{\beta}\right) f(q_z,q_x^2+q_y^2),
\end{equation}
where ${\bf n}$ is the unit vector normal to the shock front.
Since we observe along the $x$ axis, particles moving along this
direction are the main contributors to the emission, so that ${\bf
qv}=q_xv$ in the argument of $\delta$-function. Thus, we obtain:
\begin{equation}
\label{Int_q_1}
       \int  d{\bf q}
  \delta(\omega'+q_xv) f({\bf q})=\frac{1}{v}\int  dq_y
   \bar{f}( q_y^2 +(\omega'/v)^2),
\end{equation}
where $\bar{f}( q_x^2 + q_y^2)=\int f({\bf q}) dq_z$. The
correlation function $f({\bf q})$ has a peak around
$q_m=2\pi/l_{cor}$. Assuming the function $\bar{f}$ keeps this
property, which is typically the case, then at small frequencies
($\omega' \ll \omega_0$) integral (\ref{Int_q_1}) does not depend
on frequency, and the substitution of it into (\ref{I_perp_1})
gives rise to a frequency-independent spectrum
$dI_{\omega}/d\omega \propto \omega^0$, which is flatter (not
steeper) than the asymptotic spectrum of synchrotron radiation.

Thus, this flat low-frequency spectrum seems to be rather common
in the presence of small-scale random magnetic or electric fields.
However, in some specific models of the random field, the
low-frequency spectrum can deviate from the flat spectrum.

As an example of this latter point, consider a (somewhat
artificial) factorized correlation function $
 f({\bf q})=f_1(q_{\|})f_2({\bf q}_{\bot})$,
where $q_{\|}$ and ${\bf q}_{\bot}$ stand for components parallel
and transverse to the particle velocity (note that a special model
of the field, which is  uniform in the plane transverse to the
particle velocity, when $f({\bf q})=f_1(q_{\|})\delta({\bf
q}_{\bot})$, is absorbed by this particular case). Then,
integrations over $dq_{\|}$ and $d{\bf q}_{\bot}$ can be performed
independently, so that
\begin{equation}
\label{Int_q_3}
       \int  d{\bf q}
  \delta(\omega'+q_{\|}v) f({\bf q})=\frac{1}{v} f_1(\omega'/v) \int d{\bf
q}_{\bot}
   f_2({\bf q}_{\bot}),
\end{equation}
and the low-frequency asymptote of the radiation spectrum is
ultimately specified by the behavior of $f_1(q_{\|})$ at $q_{\|}
\ll q_m$ (e.g., $f_1(\omega'/v) \propto (\omega'/v)^2$ for
correlation function like (\ref{power_law_spectr})), and can
easily be steeper than $\omega^0$ spectrum with the steepest
possible asymptote $\omega^1$ (ignoring non-perturbative effects
and wave dispersion).

To see this, consider an extreme case of a small-scale field,
namely, one comprising a single spatial harmonic $ f({\bf
q})=\delta({\bf q}-{\bf q}_m)$, which evidently provides the
narrowest possible radiation spectrum. All integrations are
extremely easy in this case, in particular,
\begin{equation}
\label{Int_q_4}
       \int  d{\bf q}
  \delta(\omega'+{\bf qv}) \delta({\bf q}-{\bf q}_m)=\delta(\omega'+{\bf
q}_m{\bf v}),
\end{equation}
giving rise to the linear low-frequency asymptote
$dI_{\omega}/d\omega \propto \omega^1$, in agreement  with the
estimate of \cite{Medvedev_2000}. A similar kind of regular (but
small-scale) acceleration takes place in undulators or in the case
of so-called small-pitch-angle radiation
\citep{Epstein_1973,Epstein_Petrosian_1973}, resulting in a
similar radiation spectrum. However, in the general case of a {\it
stochastic} magnetic field, the radiation spectrum deviates
strongly from this simpler case.

\section{Non-perturbative approach}

It should be emphasized, that the applicability of the
perturbation theory developed above is itself rather limited.
Indeed, even if the deflection of the particle is small during the
time needed to pass through a single correlation cell of the
random field, it is not necessarily small along the coherence
length of the emission, since multiple scattering of the particle
by several successive magnetic inhomogeneities can easily provide
large enough deflections due to angular diffusion to render
perturbation theory inapplicable.

\begin{figure}
\epsscale{0.85} \plotone{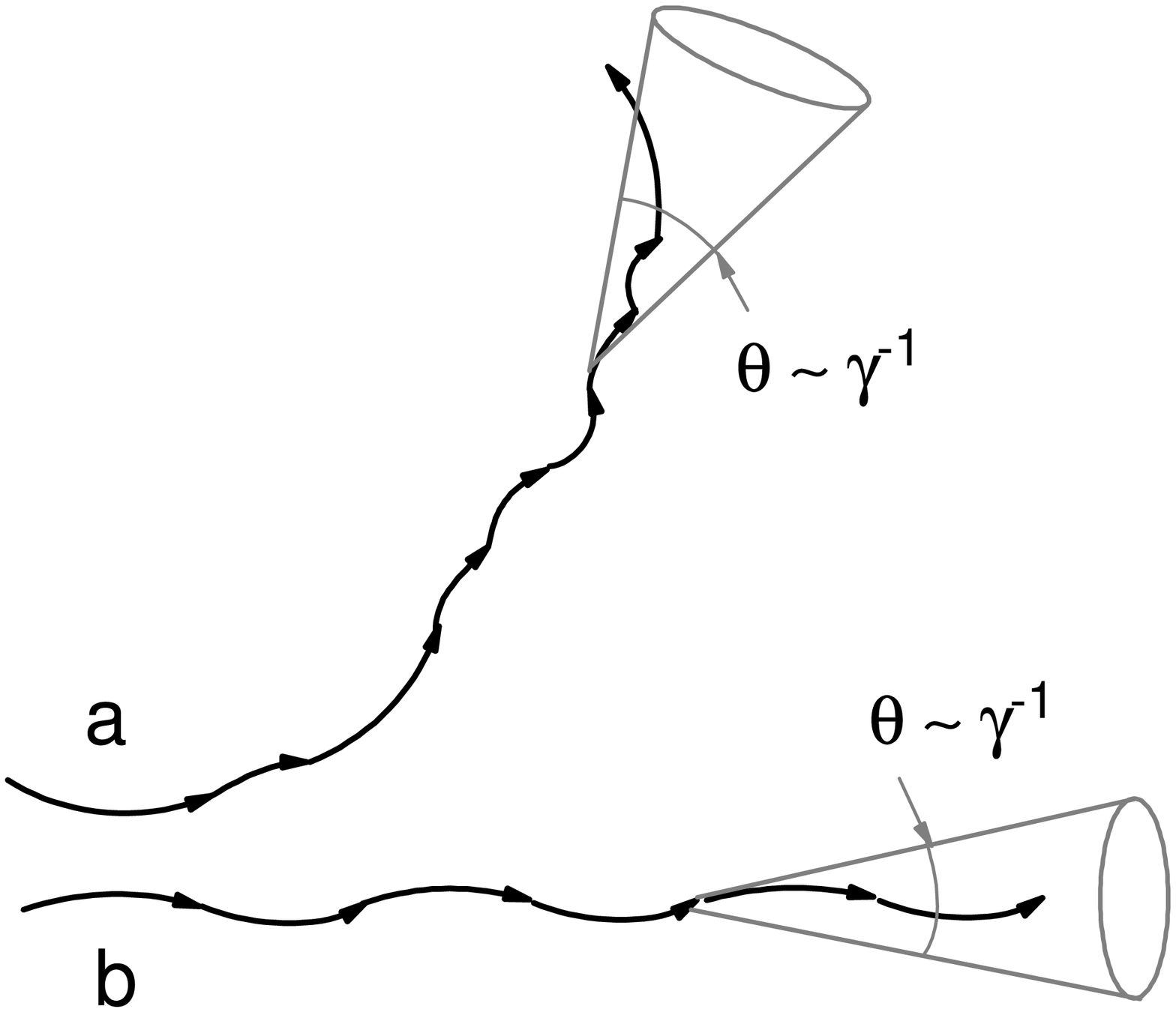} \figcaption{\small Random walk of
the particle in the random field (a) vs jitter in small-scale
regular field (b). In the former case the mean square of the
deflection angle $\theta_c$ increases with time as $\theta_c^2
\propto t$, while in the later case it remains small along the
entire trajectory. As a result, the radiation spectra deviate
strongly from each other for these two cases.
 \label{RandomWalk}}
\end{figure}

Thus, we explicitly have to  consider the applicability of
perturbation theory to the emission of fast particles moving in
the small-scale random fields. The coherence length $l_c(\omega)
\sim c\gamma_*^2/\omega$ of emission by relativistic particle with
the Lorentz-factor $\gamma$ at a frequency $\omega$ decreases as
frequency increases, so the approximation of the rectilinear
motion is eventually valid at sufficiently large frequencies.
However, at lower frequencies the coherence length may be larger
than the correlation length of the random field, and this
necessarily will be the case at low enough frequencies if one
neglects the effect of the wave dispersion in the plasma. Thus,
the particle trajectory traverses several correlation lengths of
the random field to emit this low frequency, and its trajectory
random walks due to uncorrelated scattering by successive magnetic
inhomogeneities, as depicted in fig. \ref{RandomWalk}. This
angular diffusion will clearly affect the radiation spectrum if
the mean angle of the particle deflection $\theta_c$
\emph{accumulated} along the coherence length $l_c$ exceeds the
beaming angle of the emission $\gamma^{-1}$.

To estimate the deflection angle $\theta_c$, adopt that the random
field consists of cells with the characteristic scale $l_0$ and
rms field value $\left< B_{st}^2\right>^{1/2}$. Inside each cell,
the electron velocity rotates by the angle $\theta_0 \sim
(\omega_{st}/\gamma) (l_0/c) \sim \omega_{st}/(\omega_0\gamma) $,
where $\omega_{st}=e\left< B_{st}^2\right>^{1/2}/mc$ and
$\omega_{0}=2\pi c/l_0$. Since the deflections are produced by
successive \emph{uncorrelated} cells of the random field, the mean
square of the deflection angle after traversing $N$ cells will be
$\theta_c^2 = \theta_0^2 N$, where the number of the cells is
specified by the ratio of the coherence length $l_c$ to the
correlation length $l_0$ of the random field, $N=l_c/l_0$.
Therefore, when the particle passes the length $l_c(\omega)$
required to produce the emission at a frequency $\omega$, its
characteristic deflection angle is $\theta_c^2 \simeq
\omega_{st}^2/(\omega_0 \omega)$. This angular diffusion will
strongly affect the emission if $\theta_c \gsim 1/\gamma$, i.e.,
at
\begin{equation}
\label{cond_pert_2}
 \omega \lsim  \frac{\omega_{st}^2}{\omega_0} \gamma^2,
\end{equation}
which occurs in the range of the low-frequency asymptotic limit
discussed above.

\begin{figure}
\epsscale{0.85} \plotone{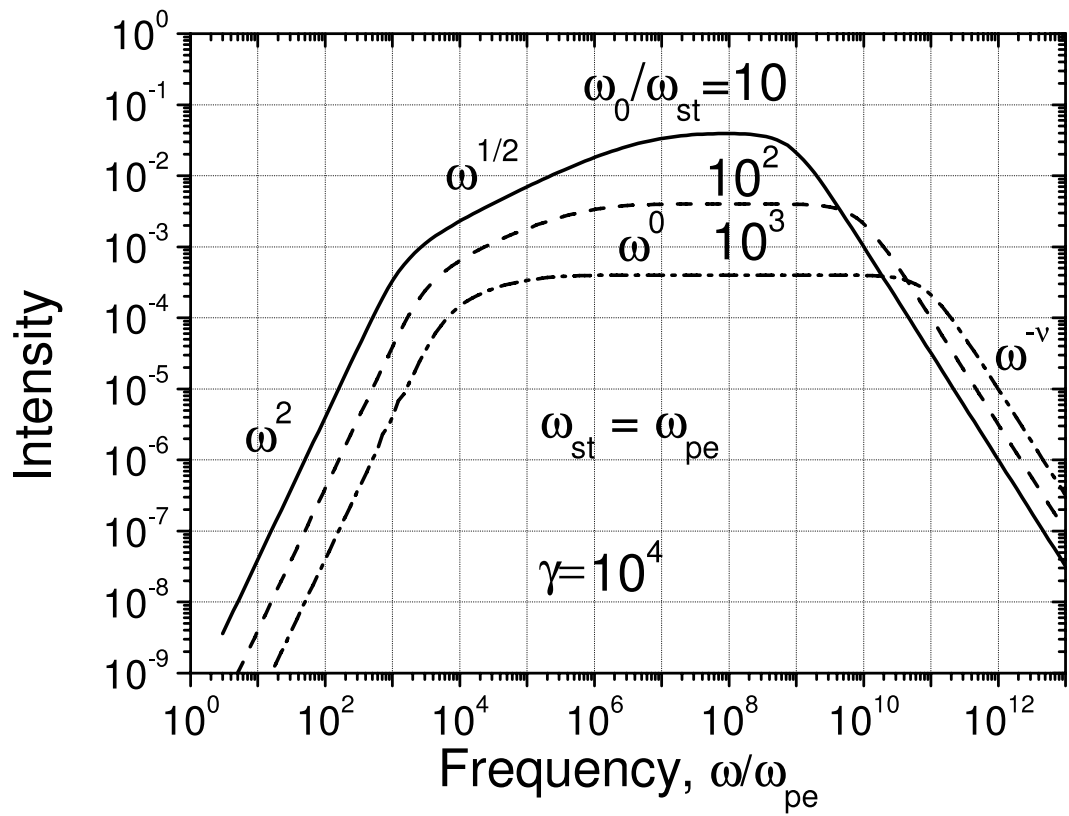} \figcaption{\small Radiation
spectra produced by a relativistic particle  with
 $\gamma=10^4$ in a small-scale random magnetic field. The
 spectral asymptote $\omega^{1/2}$ can only be obtained within the
 non-perturbative theory.
If  $\omega_0$ is large enough (e.g., $\omega_0/\omega_{st}=10^3$
in the figure) the spectral region provided by multiple
scattering, $\omega^{1/2}$, disappears, and the whole spectrum is
the same as within the perturbation theory. \label{spectra}}
\end{figure}

The non-perturbative version of the theory is not discussed in any
detail here since it is published elsewhere
\citep{Topt_Fl_1987a,Topt_Fl_1987b,Topt_etal_1987,Fl_2005b}. The
full radiation spectrum (also including the effect of a regular
magnetic field, if present) is given by Eq. (35) of
\citep{Topt_Fl_1987a}. Fig. \ref{spectra} presents radiation
spectra (calculated with the use of the non-perturbative formulae
(35) of \citep{Topt_Fl_1987a}, or, equivalently, Eq. (30) of
\citep{Fl_2005b}) for a single relativistic electron with a
Lorentz-factor $\gamma=10^4$ moving in the presence of fluctuating
small-scale fields (with different ratios $\omega_{st}/\omega_{0}=
10^{-1}, 10^{-2}, 10^{-3}$, the regular magnetic field is assumed
to be very weak, the distribution of random field is isotropic
(like (\ref{power_law_spectr})), and $\omega_{pe}=\omega_{st}$.
Although the combination of parameters adopted to produce Fig.
\ref{spectra} does not span all the regimes possibly relevant to
the GRB case (see also the review paper \cite{Fl_2005b}), it shows
the variety of asymptotes in one plot and displays spectra both
affected and unaffected by the non-perturbative effects, which
should be sufficient for illustration purpose.

The solid curve corresponds to the parameters similar to those
adopted by \cite{Medvedev_2000} for the GRB sources
($\omega_{0}=10\omega_{st}=10\omega_{pe}$). Although the flat
region obtained already within the perturbation approach exists,
it is easy to see a region of the radiation spectrum
$dI_{\omega}/d\omega \propto \omega^{1/2}$, provided by the
multiple scattering, which occupies the frequency range
\begin{equation}
\label{cond_3_case3}
\omega_{pe}(\omega_{pe}\omega_{0}\gamma^2/\omega_{st}^2)^{1/3} \ll
\omega \ll \frac{\omega_{st}^2}{\omega_{0}} \gamma^2.
\end{equation}
Note, that the left bound of this inequality is below the limit
$\omega_{pe} \gamma$ imposed by the effect of the wave dispersion
in the perturbative treatment. Stated another way, in the
non-perturbative regime the wave dispersion comes into play at
lower frequencies than expected based on the perturbative
treatment. This region of the spectrum (missing within the
perturbation theory) exists for any relativistic particle with
\begin{equation}
\label{cond_4_case3}
 \gamma \gsim \frac{\omega_{pe}\omega_0}{\omega_{st}^2},
\end{equation}
i.e., for $\gamma \gsim 10$ for the adopted parameters
$\omega_{0}=10\omega_{st}=10\omega_{pe}$, which turns out to be
important for most (if not all) available relativistic electrons
in the GRB source.

On the contrary, for smaller-scale fields, dashed and dash-dotted
curves, the angular diffusion becomes less and less important,
which justifies the applicability of the perturbation theory for
corresponding conditions. At even lower frequencies the spectrum
falls as $dI_{\omega}/d\omega \propto \omega^{2}$ due to the
effect of wave dispersion in the plasma. The high-frequency end of
the spectrum is described by the power-law $dI_{\omega}/d\omega
\propto \omega^{-\nu}$, i.e., mimics the shape of the assumed
spatial correlation function. Both these asymptotes are present in
the perturbative version of the theory.

\section{Implications for interpreting emission from GRBs}

Typically, the GRB spectrum can be well fitted with a
two-component Band model \citep{Band_etal_1993}, which represents
the observed spectrum as a sum of low-energy component,
$N(E_{\gamma}) \propto E_{\gamma}^{\alpha}\exp(-E_{\gamma}/E_0)$,
and high-energy component, $N(E_{\gamma}) \propto
E_{\gamma}^{\beta}$ with $\alpha
> \beta$. The distribution of the low energy spectral indices $\alpha$ of GRBs
\citep{Preece_etal_2000}, which is a bell-shaped curve with the
peak at  $\alpha \approx -1$ with FWHM about 1, and reaches values
$\alpha \approx 1$, is a challenge for available emission
mechanisms \citep{Baring_Braby_2004,Piran_2005}.

Indeed, optically thin  synchrotron emission can only produce low
energy indices $\alpha < -2/3$ which are incompatible with about
25$\%$ of the spectra. Inverse Compton emission (as well as a few
other more exotic processes) seems to be more promising since it
can produce any negative spectral index $\alpha < 0$, which is
compatible with roughly  98$\%$ of the spectra. The optically
thick synchrotron model
\citep{Baring_Braby_2004,Lloyd_Petrosian_2000} might accommodate
the rest 2$\%$ of the positive $\alpha$.

However, it is  remarkable that none of these bounding values
($\alpha = -2/3,~0,~1.5$) has any imprint on the distribution: it
is equally smooth at $\alpha \approx -2/3$ or $\alpha \approx 0$
and does not even reach values $\alpha \approx 1.5$.

By comparison, the GRB low-energy-spectral-index distribution
appears to be a natural outcome of diffusive synchrotron emission.
Leaving the detailed analysis and discussion of some specific
issues \citep{Piran_2005} to a future study, we note that the main
low-frequency asymptote, $\propto \omega^0$, corresponds to the
peak of the distribution, which suggests plausible interpretation
why most of the low energy indices fall around the value $\alpha =
-1$. Then, for lower frequencies, this asymptote gradually gives
way to the asymptote $\propto \omega^{1/2}$, related to the
multiple scattering effect, which is compatible with about 90$\%$
of the spectra. Therefore, the part of the diffusive synchrotron
radiation spectrum including these two asymptotes is capable of
explaining the whole range of the ``typical'' low energy spectral
indices (which we define as a confidence interval around the mean
value). The remaining 10$\%$ of the spectra might be treated as
``untypical'', probably requiring some extreme combination of
parameters or some special structure of the random field. In any
case, they are compatible with the transition to the
lowest-frequency asymptote, $\propto \omega^2$. It is especially
interesting that the latter asymptote corresponds well to the
presence of a secondary (weak but significant) peak in the
distribution at about $\alpha = 1$.

In contrast, the high-frequency asymptote, $\propto
\omega^{-\nu}$, represents the softest possible high-frequency
spectrum in the presence of small-scale random magnetic fields.
For example, if $\nu=1.5$, than the softest power spectral index
is $1.5$ and the softest {\it photon} spectral index is $2.5$
independently of the relativistic electron distribution over
energy (evidently, the spectrum can be harder than this for hard
enough electron distributions).

\section{Summary and conclusions}

Motion of a charged particle in the presence of random fields
generally represents a kind of spatial and angular diffusion,
therefore, we refer to the emission by this particle as diffusive
synchrotron radiation. This paper presents a special case of the
theory of diffusive synchrotron radiation produced by relativistic
electron deflections on a small-scale random magnetic field in the
absence of the regular magnetic field. This theory includes both
the simplified perturbation approach  and the non-perturbative
treatment, which turns out to be of primary importance in the
radiation spectrum formation.

The obtained spectrum of diffusive synchrotron radiation
substantially expands  the``jitter" radiation spectrum suggested
by \cite{Medvedev_2000} based on semi-quantitative evaluations.
The derived spectrum is composed of several power-law asymptotic
regimes smoothly giving way to each other as the emission
frequency changes, which is apparently consistent with the
observed distribution of the low energy spectral indices in GRBs.

We conclude that

$\bullet $ diffusive synchrotron radiation has a very general
nature and will be observed from any source capable of producing
small-scale random magnetic/electric fields at a sufficient level,
like active galactic nuclei, extragalactic jets, hot spots in the
radio galaxies, galactic sources with a strong energy release, as
well as solar flares,

$\bullet $ a low-frequency spectrum, $dI_{\omega}/d\omega \propto
\omega^{1}$, valid in the presence of {\it ordered} small-scale
magnetic field fluctuations, does not  occur in the general case
of small-scale \emph{random} magnetic field fluctuations,

$\bullet $ diffusive synchrotron radiation arising from the
scattering of fast electrons on small-scale {\it random} magnetic
or/and electric fields produces a broad variety of low-frequency
spectral  asymptotes -- from $dI_{\omega}/d\omega \propto
\omega^0$ to $\propto \omega^2$ -- sufficient to interpret the
entire range of low energy spectral indices observed from GRB
sources, while the high-frequency spectrum $dI_{\omega}/d\omega
\propto \omega^{-\nu}$ may affect the corresponding high energy
spectral index distribution.

\acknowledgments The National Radio Astronomy Observatory is a
facility of the National Science Foundation operated under
cooperative agreement by Associated Universities, Inc. This work
was supported in part by  the Russian Foundation for Basic
Research, grants No.03-02-17218, 04-02-39029. I am strongly
grateful to T.S. Bastian for his numerous comments to the paper
and D.A. Frail for discussion of the topic.

\begin{thebibliography}{}
\bibitem[Band et al. (1993)]{Band_etal_1993}
        Band, D., Matteson, J., Ford, L. et al. 1993, \apj, 413, 281


\bibitem[Baring \& Braby(2004)]{Baring_Braby_2004}
        Baring, M.G., \& Braby, M.L. 2004, \apj, 613, 460

\bibitem[Epstein(1973)]{Epstein_1973}
        Epstein, R.I. 1973, \apj, 183, 593

\bibitem[Epstein \& Petrosian(1973)]{Epstein_Petrosian_1973}
        Epstein, R.I., \& Petrosian, V. 1973, \apj, 183, 611

\bibitem[Fleishman (2006)]{Fl_2005b}
        Fleishman G.D., 2006, in "Geospace Electromagnetic Waves
        and Radiation",  Eds. - J.W.Labelle \& R.A.Treumann,
        Chpt. 4, pp. 83-102, Lect. Notes in Phys., V.687
        (Springer-Verlag: Berlin-Heidelberg-New York),   preprint
        astro-ph/0510317

\bibitem[Jaroshek et al. (2004)]{Jaroshek_etal_2004}
        Jaroshek, C.H., Lesch, H., \& Treumann, R.A. 2004, \apj, 616, 1065

\bibitem[Jaroshek et al. (2005)]{Jaroshek_etal_2005}
        Jaroshek, C.H., Lesch, H., \& Treumann, R.A. 2005, \apj,
        618, 822

\bibitem[Hededal \& Nishikawa  (2005)]{Hededal_Nishikawa_2005}
        Hededal, C.B., \& Nishikawa, K.-I. 2005,
        \apjl, 623, L89

\bibitem[Kazimura et al. (1998)]{Kazimura_1998}
        Kazimura, Y., Sakai, J.I., Neubert, T., \& Bulanov, S.V. 1998, \apj,
498, L183

\bibitem[Landau \& Lifshitz(1971)]{LL_2}
        Landau, L.D., \& Lifshitz, E.M. 1971 The classical theory of fields
(Oxford: Pergamon Press)

\bibitem[Lloyd \& Petrosian(2000)]{Lloyd_Petrosian_2000}
       Lloyd, N.M., \& Petrosian, V. 2000, \apj, 543, 722


\bibitem[Medvedev(2000)]{Medvedev_2000}
        Medvedev, M. V. 2000, \apj, 540, 704

\bibitem[Medvedev \& Loeb(1999)]{Medvedev_Loeb_1999}
        Medvedev, M.V., \& Loeb, A. 1999, \apj, 526, 697


\bibitem[M\'esz\'aros(2002)]{Meszaros_2002}
        M\'esz\'aros, P. \araa,  40, 137 (2002) %ARA\&A

\bibitem[Nikolaev \& Tsytovich(1979)]{Nik_Tsyt_1979}
        Nikolaev, Iu.A., \&  Tsytovich, V.N. 1979, Phys. Scripta, 20, 665

\bibitem[Nishikawa et al.\ (2003)]{Nishikawa_etal_2003}
        Nishikawa, K.-I., Hardee, P., Richardson, G., Preece, R., Sol, H., \&
        Fishman, G.J. 2003, \apj, 595, 555

\bibitem[Nishikawa et al.\ (2005)]{Nishikawa_etal_2005}
        Nishikawa, K.-I., Hardee, P., Richardson, G., Preece, R., Sol, H., \&
        Fishman, G.J. 2005,  \apj, 622, 927

\bibitem[Piran (2004)]{Piran_2005}
        Piran, T. 2004, Rev. Mod. Phys., 76, 1143


\bibitem[Preece et al. (2000)]{Preece_etal_2000}
        Preece, R.D., Briggs, M.S., Mallozzi, R.S., Pendleton, G.N.,
        Paciesas, W.S., \& Band, D.L.  2000, ApJS, 126, 19

\bibitem[Toptygin (1985)]{Toptygin_83}
        Toptygin, I.N. 1985, Cosmic rays in interplanetary magnetic fields
        (Dordrecht, D. Reidel)

\bibitem[Toptygin \& Fleishman (1987a)]{Topt_Fl_1987a}
        Toptygin, I.N., \& Fleishman, G.D. 1987a, \apss, 132, 213

\bibitem[Toptygin \& Fleishman (1987b)]{Topt_Fl_1987b}
        Toptygin, I.N., \& Fleishman, G.D. 1987b, Radiophys.
        \& Qant. Electr., 30, 551

\bibitem[Toptygin et al. (1987)]{Topt_etal_1987}
        Toptygin, I.N., Fleishman, G.D., \& Kleiner, D.V. 1987, Radiophys.
        \& Quant. Electr., 30, 334

%\bibitem[Tsytovich \& Chikhachev(1969)]{Tsytovich_Chikhachev_1969}
%        Tsytovich, V.N. \& Chikhachev, A.S. 1969, Sov. Astron.,
%        13, 385

\bibitem[Weibel (1959)]{Weibel_1959}
        Weibel, E. S. 1959, \prl, 2, 83 %Phys. Rev. Lett.

\end{thebibliography}
\end{document}